\documentclass[journal,10pt]{IEEEtran}

\IEEEoverridecommandlockouts
\usepackage{amsbsy}
\usepackage{amsthm}
\usepackage{amssymb}
\usepackage{amsmath}
\usepackage{multirow}
\usepackage{epsfig}
\usepackage{subfigure}
\usepackage{graphics}
\usepackage{multicol}

\usepackage{epstopdf}
\usepackage{balance}
\usepackage{xcolor}
\usepackage{cite}
\usepackage{color}
\usepackage{algorithm}
\usepackage{algorithmic}
\usepackage{textcomp}
\usepackage{subfigure}
\usepackage{subfloat}
\usepackage{booktabs}
\usepackage{bm}
\usepackage{filecontents}
\usepackage{upgreek}
\usepackage{setspace}
\newtheorem{Def}{Definition}
\newtheorem{Them}{Theorem}

\begin{document}

\title{\huge{Rethinking Estimation Rate for Wireless Sensing: \\ A Rate-Distortion Perspective}}

\author{Fuwang Dong,~\IEEEmembership{Member,~IEEE}, Fan Liu,~\IEEEmembership{Member,~IEEE}, Shihang Lu, Yifeng Xiong,~\IEEEmembership{Member,~IEEE}  \vspace{-2em}}
{}
\maketitle

\begin{abstract} Wireless sensing has been recognized as a key enabling technology for numerous emerging applications. For decades, the sensing performance was mostly evaluated from a reliability perspective, with the efficiency aspect widely unexplored. Motivated from both backgrounds of rate-distortion theory and optimal sensing waveform design, a novel efficiency metric, namely, the sensing estimation rate (SER), is defined to unify the information- and estimation- theoretic perspectives of wireless sensing. Specifically, the active sensing process is characterized as a virtual lossy data transmission through non-cooperative joint source-channel coding. The bounds of SER are analyzed based on the data processing inequality, followed by a detailed derivation of achievable bounds under the special cases of the Gaussian linear model (GLM) and semi-controllable GLM. As for the intractable non-linear model, a computable upper bound is also given in terms of the Bayesian Cram\'er-Rao bound (BCRB). Finally, we show the rationality and effectiveness of the SER defined by comparing to the related works.          
\end{abstract}


\IEEEpeerreviewmaketitle

\section{Introduction}\label{Introduction}

\subsection{Motivation and Related Works}
Sensing tasks can be roughly classified into three categories, \textit{detection}, \textit{estimation}, and \textit{recognition}, which are essentially deducing the desired parameters/features from the collected signals/data with respect to the targets. Most of the sensing performance metrics are reliability metrics, such as detection probability and mean square error (MSE), etc, with the effciency aspect of sensing widely unexplored \cite{liu2022integrated}. In light of the recent progress of integrated sensing and communications (ISAC) techniques \cite{liu2022integrated}, evaluating the sensing performance by a well-defined information quantity consistent with the communication metrics, is envisioned as an essential methodology in revealing the fundamental limits and performance trade-off of ISAC, which may provide profound insights into system design and resource allocation \cite{liu2022survey}.     

The rate-distortion theory, which initially characterizes the minimal information rate required to achieve a preset distortion in lossy data transmission \cite{thomas2006elements}, bridges the quantities between information and estimation. Based on this principle, the mutual information (MI) based radar waveform design has been presented \cite{bell1993information,tang2018spectrally}, where the MI between the observations and the target impulse response is maximized under the Gaussian linear model (GLM). In \cite{yang2007mimo}, the authors show that maximizing MI and minimizing minimum MSE (MMSE) through the waveform design lead to the same water-filling solution. Nevertheless, the operational meaning of the sensing MI, as well as its connection with various commonly employed sensing metrics still remain unclear, which limits the applications of the MI-based methods for sensing or ISAC systems.           

Recently, the capacity-distortion-cost trade-off of state-dependent memoryless channel was studied in \cite{ahmadipour2022information}, where the channel state parameters are simultaneously estimated through the backscattered signals while communicating with the users. Although the considered scenario is common in wireless communications, the yielded capacity-distortion trade-off is more applicable to the case where the sensing and communication channels are $\textit{strongly correlated}$, e.g., the communication user is also a target to be sensed. As a consequence, it is unable to address more general issue where there is only partial correlation between two channels. More related to this work, a radar estimation information rate has been defined in \cite{chiriyath2015inner} for target range estimation. Nevertheless, its operational meaning was not fully investigated in a rigorous manner. 

\subsection{Our Contributions}
To address the above issues, in this paper, we define the sensing estimation rate (SER) for the general sensing model from the perspective of rate-distortion theory and waveform design, where the sensing process is analogized as a virtual non-cooperative communication process. The SER is expected to be a stepping stone towards unifying the information-theoretic and estimation-theoretic metrics, thereby providing an analytical framework for depicting the performance gain of ISAC systems. The contributions are summarized as follows.       

\begin{itemize}
	\item Based on the rate-distortion theory, sensing process is characterized by an equivalent virtual lossy data transmission process. Namely, a virtual user ``transmits'' information of parameters through joint source-channel coding under the constraint of ``channel capacity''.
	
	\item The SER is defined as the minimal information rate required to achieve the lower bound of estimation error through the optimal waveform design, followed by the bounds analysis in terms of data processing inequality.  
	
	\item The achievable bounds are derived for the special cases of GLM and semi-controllable GLM. Furthermore, a computable upper bound of SER for the nonlinear Gaussian model is also given.  
\end{itemize} 

Under the proposed framework, we conclude that: (1) The results in \cite{yang2007mimo} is a weak version of SER; (2) The estimation information rate defined in \cite{chiriyath2015inner} is a special scalar case for SER; (3) The MI-based sensing system designs may be endowed with an operational estimation-theoretic meaning.

\begin{figure}[!t]
	\centering
	\includegraphics[width=3.5in]{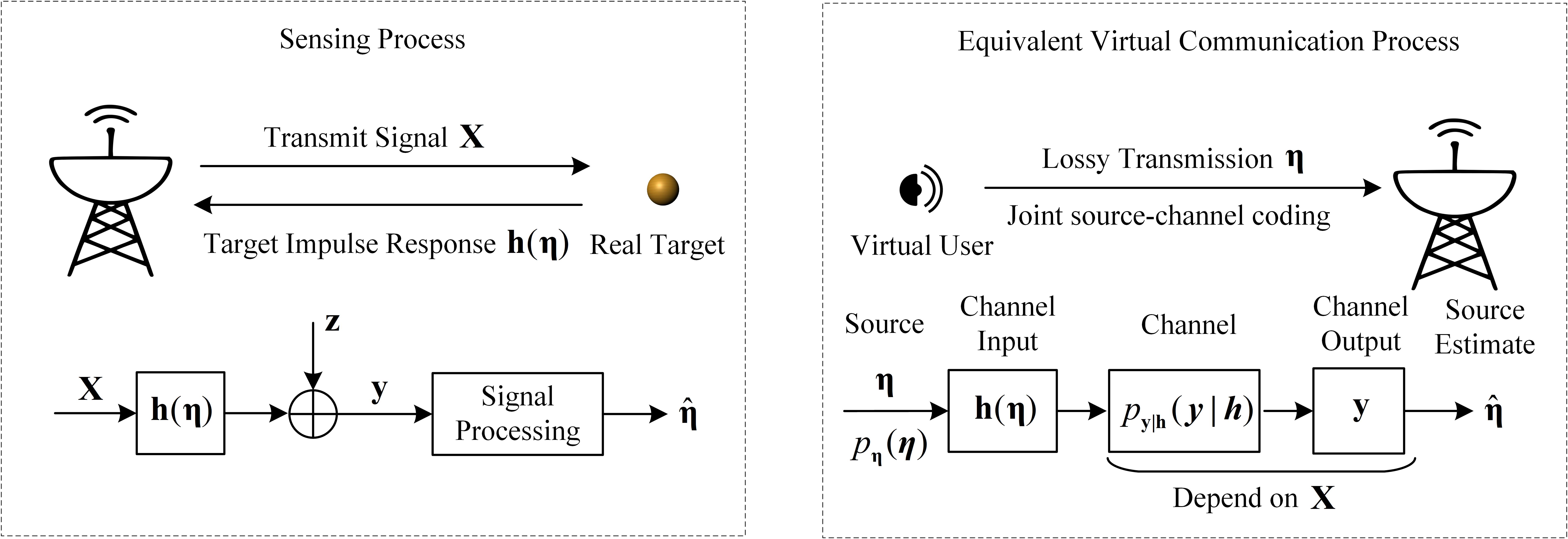}
	\caption{Active sensing as non-cooperative joint source-channel coding.}
	\label{rate_distortion_system}
\end{figure}

\section{The Definition of Estimation Rate}\label{SystemModel}

\subsection{The Sensing System Model}
Let us consider a general sensing model under Gaussian channel, where the received signal can be expressed as
\begin{equation}\label{M1}
	\textbf{Y}=\textbf{H}(\bm{\upeta})\tilde{\textbf{X}}+\textbf{Z},
\end{equation}
where $\tilde{\textbf{X}} \in \mathbb{C}^{M_t \times T} $ represents the matrix of transmitted waveform, $T$ is the number of discrete samples, $\textbf{Z} \in \mathbb{C}^{M_r \times T}$ denotes the additive noise which follows an independently and identically distributed (i.i.d.) circularly symmetric complex Gaussian distribution, $M_t$ and $M_r$ are the numbers of transmitting and receiving antennas, and $\textbf{H}(\bm{\upeta}) \in \mathbb{C}^{M_r \times M_t}$ is the sensing channel matrix with respect to the latent parameters $\bm{\upeta} \in \mathbb{R}^{K}$. The sensing task aims at recovering $\bm{\upeta}$ from the noisy observation $\textbf{Y}$ via a well-designed estimator $\hat{\bm{\upeta}}$. Here, $\bm{\upeta}$  can be classified into several categories, which correspond to different sensing tasks.
\begin{itemize}
	\item \textit{Target Parameter Estimation} ($\bm{\upeta} \in \mathbb{R}^{K}$): In this case, $\bm{\upeta}$ refers to the physical parameters of targets, e.g., amplitude, delay, angle, Doppler, etc.
	
	\item \textit{Channel Estimation} ($\bm{\upeta}=\text{vec}({\textbf{H}})$): This case also corresponds to \textit{linear spectrum estimation}, where the receiver only requires the knowledge of channel state information (CSI) or channel impulse response (CIR).
	
	\item \textit{Target Detection} ($\bm{\upeta}\in \{0,1\}$): Here, $0$ and $1$ represent the absence and presence of the target in hypothesis testing, respectively. The estimator $\hat{\bm{\upeta}}$ is exactly the decision rule.    
\end{itemize} 

Furthermore, we highlight that (\ref{M1}) may describe a variety of sensing scenarios with different forms of $\textbf{H}(\bm{\upeta})$. 
Specifically, we take the following two commonly applied sensing models as examples.

$\bullet$ For parameter estimation of targets using a MIMO radar equipped with co-located antennas, the sensing channel matrix $\textbf{H}(\bm{\upeta})$ has the following form \cite{li2007range}
\begin{equation}
	\textbf{H}=\sum_{i=1}^L \alpha_i \textbf{a}(\theta_i)\textbf{b}^H(\theta_i),
\end{equation}
where $L$ is the number of the targets, $\alpha_i$ and $\theta_i$ are the complex reflection coefficient and the angle of the $i$-th target. $\textbf{a}(\theta)$ and $\textbf{b}(\theta)$ are the receiving and transmitting steering vectors, respectively. In this case, $\bm{\upeta}=[{\bm{\theta},\text{Re}\{\bm{\alpha}\}, \text{Im}\{\bm{\alpha}\}}]$.  

$\bullet$ For discrete-time orthogonal frequency division multiplexing (OFDM) system with $N$ subcarriers, the CIR matrix $\textbf{H}(\bm{\upeta})$ with sufficient cyclic prefix can be expressed by \cite{4600229} 
\begin{equation}
	\textbf{H}=
	{\scriptsize  \left[ {\begin{array}{*{20}{c}}
				{h(0,\tau_0)}&{0}&{\cdots}&{h(0,\tau_{L-1})}&{\cdots}&{h(0,\tau_1)}\\
				{h(1,\tau_1)}&{h(1,\tau_0)}&{0}&{\cdots}&{\cdots}&{h(0,\tau_2)}\\
				{\vdots}&{\vdots}&{\ddots}&{\vdots}&{\vdots}&{\vdots}\\
				{0}&{0}&{\cdots}&{h(N-1,\tau_{L-1})}&{\cdots}&{h(N-1,\tau_0)}
		\end{array}} 
		\right]}
\end{equation}
where $h(n,m)$ represent the multi-path time- and frequency- selective fading channel at time $n$, with the expression of 
\begin{equation}
	h(n,m)=\sum_{i=0}^L \alpha_i \delta(m-\tau_i)e^{j2\pi f_{\text{D}_i} n},
\end{equation}
where $L$ is the number of path. The parameter vector $\bm{\upeta}=[\alpha_i, \tau_i, f_{\text{D}_i}]$ represent the time-varying path gain, time delay and Doppler frequency of the $i$-th path, respectively.

In this paper, we consider $\textbf{H}$ as an unstructured matrix that contains useful information about the targets for generic sensing model, rather than restricting to a specific scenario. The vector form of (\ref{M1}) can be rewritten by
\begin{equation}\label{VecModel}
	\textbf{y}=\textbf{X}\textbf{h}+\textbf{z},
\end{equation}  
where $\textbf{X}=\tilde{\textbf{X}}^T \otimes \textbf{I}_{M_r \times M_r}$ represents the equivalent waveform matrix. $\textbf{y}=\text{vec}(\textbf{Y})$, $\textbf{h}=\text{vec}(\textbf{H})$, $\textbf{z} = \text{vec}(\textbf{Z})\sim \mathcal{CN}(0, \sigma_z^2\textbf{I}_{M_r \times T})$, $\sigma_z^2$ is the variance of noise. 

We assume that the parameters take values on a set $\mathcal{A}$ with a prior distribution $p_{\bm{\upeta}}(\bm{\eta})$. The sensing performance can be measured by the expected distortion between the estimate and the ground truth, i.e.,   
\begin{equation} {\label{Dist}} 
	D=\mathbb{E}_{\bm{\upeta}}\left[d(\bm{\upeta},\hat{\bm{\upeta}})\right],
\end{equation}  
where $d:\mathcal{A} \times \hat{\mathcal{A}} \to \mathbb{R}$ is a distance metric of estimation error. The choice of $d$ depends on the specific sensing task. For instance, $D$ is the MSE with the Euclidean distance $d=\|\bm{\upeta}-\hat{\bm{\upeta}}\|_2^2$ for parameter estimation, and is equivalent to negative detection probability with the Hamming distance $d(\bm{\upeta},\hat{\bm{\upeta}})=\bm{\upeta} \oplus \hat{\bm{\upeta}}$ for target detection.

\subsection{The Definition of SER}
In an information-theoretic sense, wireless sensing and communication are intertwined with each other as an ``odd couple''. This motivates us to characterize wireless sensing as a procedure of non-cooperative joint source-channel coding. As shown in Fig. \ref{rate_distortion_system}, the target can be regarded as a virtual user which encodes the information of $\bm{\upeta}$, and communicates it to the sensing receiver through lossy data transmission. To be specific, the random parameter $\bm{\upeta}$ is regarded as a memoryless source, $\textbf{h}(\bm{\upeta})$ and $\textbf{y}$ are the input and output of the channel $p_{\textbf{y}|\textbf{h}}$, respectively. Finally, a distorted version of $\bm{\upeta}$ may be reconstructed from the noisy data $\textbf{y}$ at the receiver.   

It is evident that $\bm{\upeta} \to \textbf{h} \to \textbf{y} \to \hat{\bm{\upeta}}$ forms a Markov chain. In such a case, the rate-distortion function characterizes the minimum number of bits required to communicate $\bm{\upeta}$ to the receiver with a preset distortion $D$ between the source (ground truth of sensing parameters) and the reconstructed information (estimate). Furthermore, we highlight that the achievable distortion is determined by both the ``capacity'' of channel $p_{\textbf{y}|\textbf{h}}$ and the transmit waveform $\textbf{X}$. Accordingly, we provide the following definition of SER to bridge the quantities between the information and estimation.   

\begin{Def}
	The Estimation Rate $R_{\rm{E}}$ for sensing systems modeled by (\ref{VecModel}), is defined as
	\begin{equation}\label{ER}
		R_E(D_L)=\mathop { \rm{min} } \limits_{P_{\hat{\bm{\upeta}}|\bm{\upeta}}:\mathbb{E}\left[d(\bm{\upeta},\hat{\bm{\upeta}})\right] = D_L} I(\bm{\upeta};\hat{\bm{\upeta}}),
	\end{equation}
	where $I(\bm{\upeta};\hat{\bm{\upeta}})$ represents the MI between $\bm{\upeta}$ and $\hat{\bm{\upeta}}$, $D_L$ is the lower bound of the achievable distortion over all controllable operations (for instance, waveform and estimator designs), i.e., 
	\begin{equation}\label{DL}
		D_L = \mathop { \rm{min} }\limits_{\hat{\bm{\upeta}} \in \hat{\mathcal{A}}, \kern 2pt \mathbf{X} \in \mathcal{X}} \kern 5pt \mathbb{E}\left[d(\bm{\upeta},\hat{\bm{\upeta}})\right].
	\end{equation}
	Here, $\mathcal{X}$ represents the feasible set of transmit waveforms. Each element in $\mathcal{X}$ satisfies $ \| \mathbf{X}\|_F^2 \le TP_T$, where $P_T$ is the power budget at transmitter. 
\end{Def}

\textbf{Remark 1:} Compared to the conventional rate-distortion function, the distortion $D_L$ in SER has a explicit physical meaning which represents the expectations of estimation error determined by the practical sensing channel $p_{\textbf{y}|\textbf{h}}$ and the transmit waveform $\mathbf{X}$, rather than an arbitrary desired distortion $D$ in lossy source coding. In other words, the SER defined can be interpreted as the minimal required information rate to communicate $\bm{\upeta}$ from the target (virtual communication transmitter) to the sensing receiver through the sensing channel, such that the minimum estimation error (distortion) can be achieved at the receiver.

\subsection{The Bounds for SER}
Note that the SER is a monotonically decreasing function of the distortion, namely, reaching a lower $D_L$ would require higher $R_E$, and vice versa. By recalling the Markov chain $\bm{\upeta} \to \textbf{h} \to \textbf{y} \to \hat{\bm{\upeta}}$, we have    
\begin{equation} \label{SourceCoding}
	R_E =\mathop { \rm{minimize} } \limits_{P_{\hat{\bm{\upeta}}|\bm{\upeta}}:\mathbb{E}\left[d(\bm{\upeta},\hat{\bm{\upeta}})\right] = D_L} I(\bm{\upeta};\hat{\bm{\upeta}}) \le I(\bm{\upeta};\hat{\bm{\upeta}}) \mathop \le \limits^{(a)} I(\textbf{y};\textbf{h}), 
\end{equation}
where $(a)$ is due to data processing inequality, and $I(\textbf{y};\textbf{h})$ is the abbreviated form of $I(\textbf{y};\textbf{h}|\textbf{X})$ without causing confusion. Furthermore, $I(\textbf{y};\textbf{h})$ can be upper bounded by the following inequality
\begin{equation} \label{ChannelCoding}
	I(\textbf{y};\textbf{h}) \mathop \le \limits^{(b)} \mathop \text{max} \limits_{\textbf{X} \in \mathcal{X}}  I(\textbf{y};\textbf{h}) \triangleq I(\textbf{X}^\star) \mathop \le \limits^{(c)} \sup_{\textbf{X} \in \mathcal{X}, p_\textbf{h}} I(\textbf{y};\textbf{h}) \triangleq C(\textbf{X}^\star),
\end{equation}
where $I(\textbf{X}^\star)$ is the MI with respect to optimal waveform $\textbf{X}^\star$ for a given channel prior $p_\textbf{h}$. $C(\textbf{X}^\star)$ is the maximum MI by maximizing over both channel prior $p_\textbf{h}$ and transmit waveform set $\mathcal{X}$. The conditions for the equality are given as follows.   
\begin{itemize}
	\item $(a)$ holds if and only if $I(\bm{\upeta};\textbf{y}|\hat{\bm{\upeta}})=0$;\footnote{In this case, we have $I(\bm{\upeta};\hat{\bm{\upeta}})=I(\textbf{y};\bm{\upeta})$ and apply the fact that $I(\textbf{y};\bm{\upeta}) = I(\textbf{y};\textbf{h})$, as $\textbf{h}$ is a deterministic function of $\bm{\upeta}$ \cite{liu2022deterministic}.}   
	
	\item $(b)$ holds by the optimal waveform design;
	
	\item $(c)$ holds when the prior distribution $p_{\textbf{h}}$ is Gaussian.  
\end{itemize}
Thus, according to the monotonicity of rate-distortion function, the distortion can be lower bounded by
\begin{equation} 
	D_L \ge D(I(\textbf{X}^\star)) \ge D(C(\textbf{X}^\star)).
\end{equation}

\textbf{Remark 2:} In the above discussions, the specific encoding and decoding methods are omitted since there is no actual coding happened. The bound derived is based on the extremum principle and the data processing inequality. This process is similar to the converse proof for the source-channel separation theorem, in the sense that (\ref{SourceCoding}) depends only on the prior (source) and the loss function, and (\ref{ChannelCoding}) depends only on the channel statistical model \cite{yp2023information}. It shows that the SER quantitatively connects the reliability metrics with the efficiency metrics by providing a lower bound $D(C(\textbf{X}^\star))$ for any achievable distortion.


\section{The Achievable Bounds for Special Cases} \label{Example}
Since it is difficult to obtain explicit expressions of SER in general, in this section, we reveal further insights into the SER metric by focusing on the \textit{quadratic Gaussian problem}. To be more specific, we consider the MSE distortion metric and Gaussian source $\bm{\upeta}$. According to the results of rate-distortion function in \cite[Th.10.3.2 \& 10.3.3]{thomas2006elements}, we have\footnote{The coefficient $\frac{1}{2}$ is vanished since the complex variable is considered.}   
\begin{equation} \label{RI}
	R_E(D_L) = \log^+ (\sigma_{\upeta}^2/D_L), 
\end{equation}
with $\sigma_{\upeta}^2$ being the variance of $\upeta$ for scalar case, and   
\begin{equation}\label{VRD}
	R_E(D_L) = \sum_{i=1}^M \log \frac{\sigma^2_{\upeta_i}}{D_i}, \kern 5pt 	D_i = \left\{
	\begin{aligned}
		\xi, \kern 2pt \text{if} \kern 2pt \xi < \sigma^2_{\upeta_i}, \\
		\sigma^2_{\upeta_i}, \kern 2pt \text{if} \kern 2pt \xi \ge \sigma^2_{\upeta_i},
	\end{aligned} \right.
\end{equation}
for $M$ independent Gaussian random variables with variance $\sigma_{\upeta_i}^2$, where $\xi$ is the inverse water-filling factor such that $\sum_{i=1}^M D_i=D_L$. In what follows, we will derive the achievable bounds for the cases of GLM, semi-controllable GLM and non-Gaussian model based on the explicit expressions of rate-distortion in (\ref{RI}) and (\ref{VRD}), and reveal the connections between the SER defined in this paper and the existing works.

\subsection{Case 1: GLM}\label{IRE}
Let us first consider the GLM model, which is typical in communication channel or radar impulse response estimation \cite{yang2007mimo,tang2018spectrally,bell1993information}. In such a case, we have $\bm{\upeta} = \textbf{h}$ with $\bm{\upeta} \sim \mathcal{CN}(\textbf{0},\bm{\Sigma}_{\bm{\upeta}})$, where $\bm{\Sigma}_{\bm{\upeta}}$ represents the covariance matrix. It is well known that the lower bound of the estimation error can be achieved by the MMSE estimator for this classical GLM, which leads to the following theorem. 
\begin{Them}
	For GLM, let $D_L = \rm{MMSE}(\mathbf{X}^\star)$, where $\mathbf{X}^\star$ is the optimal waveform that minimizes the MMSE. Under such a setting, the SER achieves its upper bound, i.e.,     
	\begin{equation}\label{T1}
		R_E(D_L)=I(\mathbf{X}^\star)=C(\mathbf{X}^\star).
	\end{equation}
\end{Them}
\textit{Proof}: The proof is relegated to Appendix \ref{AppendixA}. $\hfill\blacksquare$

\textbf{Remark 3}: It should be highlighted that (\ref{T1}) only holds for the optimal waveform. We will verify that there exists a gap between SER and MI for an arbitrary waveform $\textbf{X}$ in Section \ref{simulation}. The pioneering work \cite{yang2007mimo} has shown that minimizing MMSE and maximizing MI $I(\textbf{y};\bm{\upeta})$ lead to an identical solution for radar waveform design. This result can be seen as a weak version of \textit{Theorem 1}, since (\ref{T1}) shows a completely equivalent relations. In addition, \textit{Theorem 1} also endows the MI-based sensing system design (e.g., \cite{tang2018spectrally}) with the estimation-theoretic meaning. That is, for impulse response estimation under GLM, maximizing MI is equivalent to achieving the SER, thereby reaching the MMSE.          

\subsection{Case 2: Semi-Controllable GLM}
In this subsection, we consider the parameter estimation problem with a linear (or approximate linear) function $\textbf{h}(\bm{\upeta})=\textbf{F}\bm{\upeta}$. An example can be found in \cite{liu2020radar}, where $\textbf{F}$ is the Jacobian matrix in the Kalman filtering. Although the system model is still a GLM by treating $\textbf{X}\textbf{F}$ as a new matrix, it is quite different from the model in \ref{IRE} in terms of waveform design. Since $\textbf{F}$ is a fixed (uncontrollable) matrix related to the channel state, we refer to the model $\textbf{y}=\textbf{X}\textbf{F}\bm{\upeta}+\textbf{z}$ as the semi-controllable GLM, leading to a different waveform design strategy.
\begin{Them}
For the GLM $\mathbf{y}=\mathbf{X}\mathbf{F}\bm{\upeta}+\mathbf{z}$, $\rm{MMSE}(\tilde{\mathbf{X}}^\star)$ and $I(\mathbf{X}^\star)$ represent the minimum MMSE and maximum MI with respect to optimal waveform $\tilde{\mathbf{X}}^\star$ and $\mathbf{X}^\star$, respectively. We have
	\begin{equation}\label{T2}
		R_E(\rm{MMSE}(\tilde{\mathbf{X}}^\star)) \le I(\mathbf{X}^\star) = C(\mathbf{X}^\star),
	\end{equation}
	where the equality holds if matrix $\mathbf{F}$ has identical non-zeros eigenvalues and its right singular space equals to the eigenspace of the covariance matrix $\bm{\Sigma_{\bm{\upeta}}}$.
\end{Them}    
\textit{Proof}: The proof is relegated to Appendix \ref{AppendixB}. $\hfill\blacksquare$ 
  
{\it Theorem 2} shows that maximizing MI and minimizing MMSE are no longer equivalent under the semi-controllable GLM, if matrix $\textbf{F}$ cannot satisfy the strict conditions such that the equality holds. In other words, the SER cannot achieve the maximum MI, which implies that the optimal waveform for maximizing the MI is sub-optimal for minimizing the MMSE.

\subsection{Discussion on Non-Gaussian Model}
For non-linear $\textbf{h}(\bm{\upeta})$ with Gaussian distributed $\bm{\upeta}$, there are even no explicit expressions for MI and MMSE in general, since $\textbf{h}(\bm{\upeta})$ is not Gaussian. To this end, we consider the Bayesian Cram\'er-Rao bound (BCRB) of $\bm{\upeta}$, which provides a lower bound for the MSE of weakly unbiased estimators. Based on the result in Theorem 2, we have     
\begin{equation}\label{NonGLM}
	R_E(\text{BCRB}(\textbf{X}^\star)) \le \sum_{i=1}^M\log \left(\sqrt{\frac{\sigma_z^2\bm{\Lambda}_{\text{J}_{i}}}{\bm{\Lambda}_{\text{G}_{i}}}}\lambda-\frac{\sigma_z^2}{\bm{\Lambda}_{\text{G}_{i}}}\right)^+.
\end{equation}
The definitions of the symbols used in (\ref{NonGLM}), the detailed proof, and the conditions for equality are relegated into Appendix \ref{AppendixC}.  

\textbf{Remark 4}: The BCRB is known as a loose bound and is not asymptotically achieved by the MMSE, especially for the case where the Jacobian matrix $\textbf{F}_{\bm{\upeta}} = \partial \textbf{h}(\bm{\upeta})/\partial \bm{\upeta}$ depends on the parameter $\bm{\upeta}$. However, if the Jacobian matrix $\textbf{F}_{\bm{\upeta}}=\textbf{F}$ is independent to $\bm{\upeta}$, the BCRB is an asymptotically tight bound. 
	
By resorting to the definition in (\ref{NonGLM}), let us reconsider the non-linear scalar time-delay estimation in \cite{chiriyath2015inner} with the following received signal model  
\begin{equation}
	y(t)=\alpha x(t-\upeta)+z(t),
\end{equation} 
where the parameter $\upeta = \tau$ denote the time-delay of the reflecting signal. The BCRB of $\bm{\upeta}$ can be calculated by
\begin{equation}\label{CRB}
	\text{BCRB}_{\upeta} = (J_{\text{D}} + J_\text{P})^{-1}= (\frac{1}{\mathbb{E}_{\upeta}[\sigma^2_{\text{CRB}}]}+\frac{1}{\sigma^2_{\upeta}})^{-1},
\end{equation} 
where $\sigma^2_{\text{CRB}}=(8 \pi^2 B^2_\text{rms} \text{SNR})^{-1}$ represents the Cram\'er-Rao bound for a deterministic parameter $\eta$ and SNR is the signal to noise ratio. $B^2_\text{rms}=\int f^2 \left| {X(f)} \right|^2 df/\int \left| {X(f)} \right|^2 df$ is the effective bandwidth with $X(f)$ being the Fourier transform of the waveform $x(t)$. It is evident that we have $\mathbb{E}_{\upeta}[\sigma^2_{\text{CRB}}]=\sigma^2_{\text{CRB}}$, since $\sigma^2_{\text{CRB}}$ is independent to the true value of $\upeta$. In such a case, the BCRB is asymptotically tight and the equality in (\ref{NonGLM}) holds with scalar $\upeta$. Therefore, by combining Definition 1 and formula (\ref{RI}), the SER can be obtained by
\begin{equation}\label{BLS}
	R_E=\log \left(\frac{\sigma^2_{\upeta}}{\text{BCRB}_\upeta} \right)^+=\log \left(1+\frac{\sigma^2_{\upeta}}{\sigma^2_{\text{CRB}}}\right).
\end{equation} 

\textbf{Remark 5}: The radar estimation information rate has been defined in \cite{chiriyath2015inner} without giving a clear explanation. It should be noted that $R_{\text{E}}$ is consistent with radar estimation information rate given in \cite[formula (16)]{chiriyath2015inner} by omitting the coefficient of pulse repetition interval. However, the SER defined in this paper has a clear information-estimation relation from the perspective of rate-distortion theory and can be extended to more general scenarios.   

\begin{figure}[!t]
	\centering
	\includegraphics[width=3in]{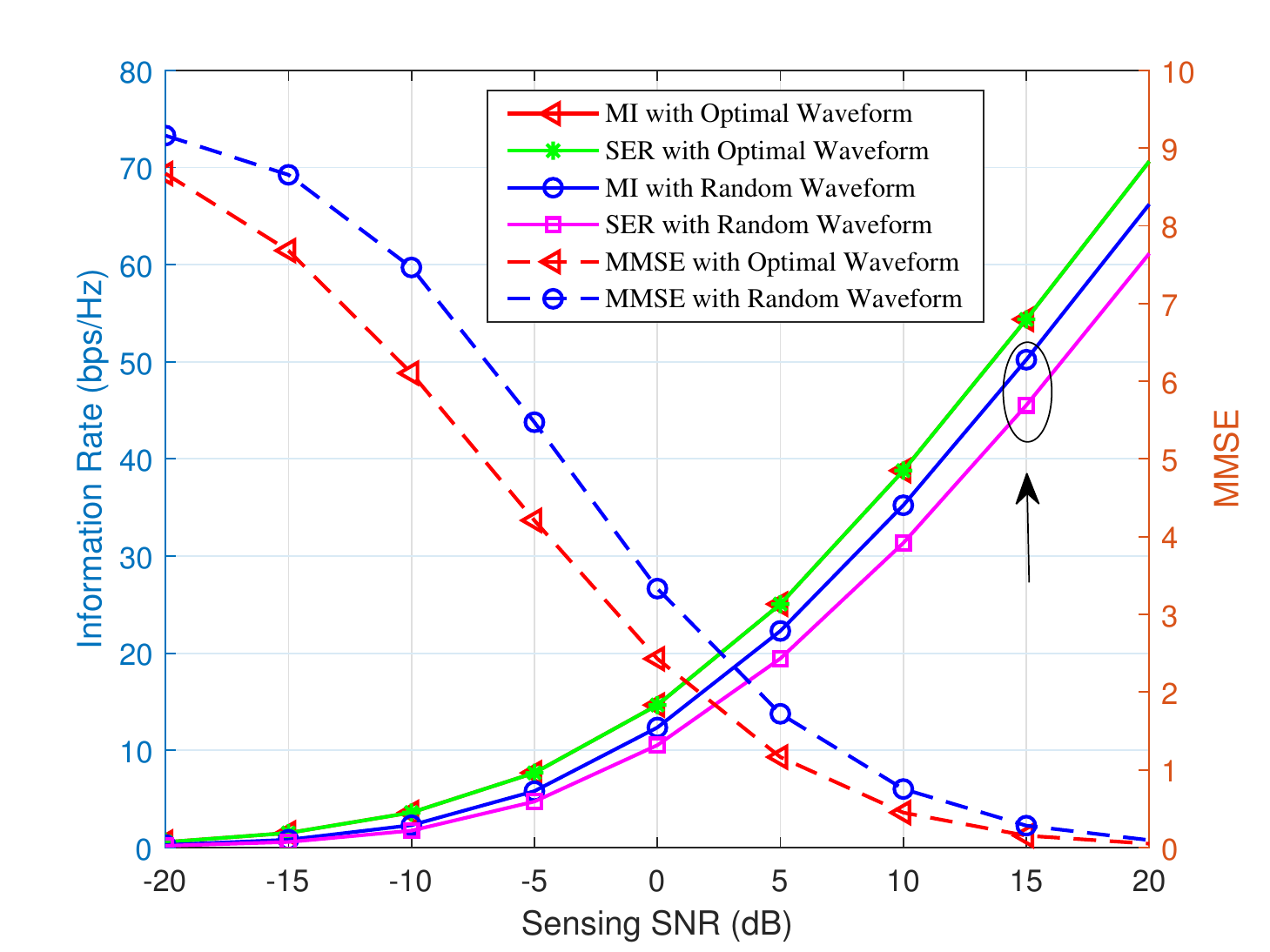}
	\caption{The curves of MI, MMSE, and SER in the GLM case.}
	\label{RandX}
\end{figure}

\begin{figure}[!t]
	\centering
	\includegraphics[width=3in]{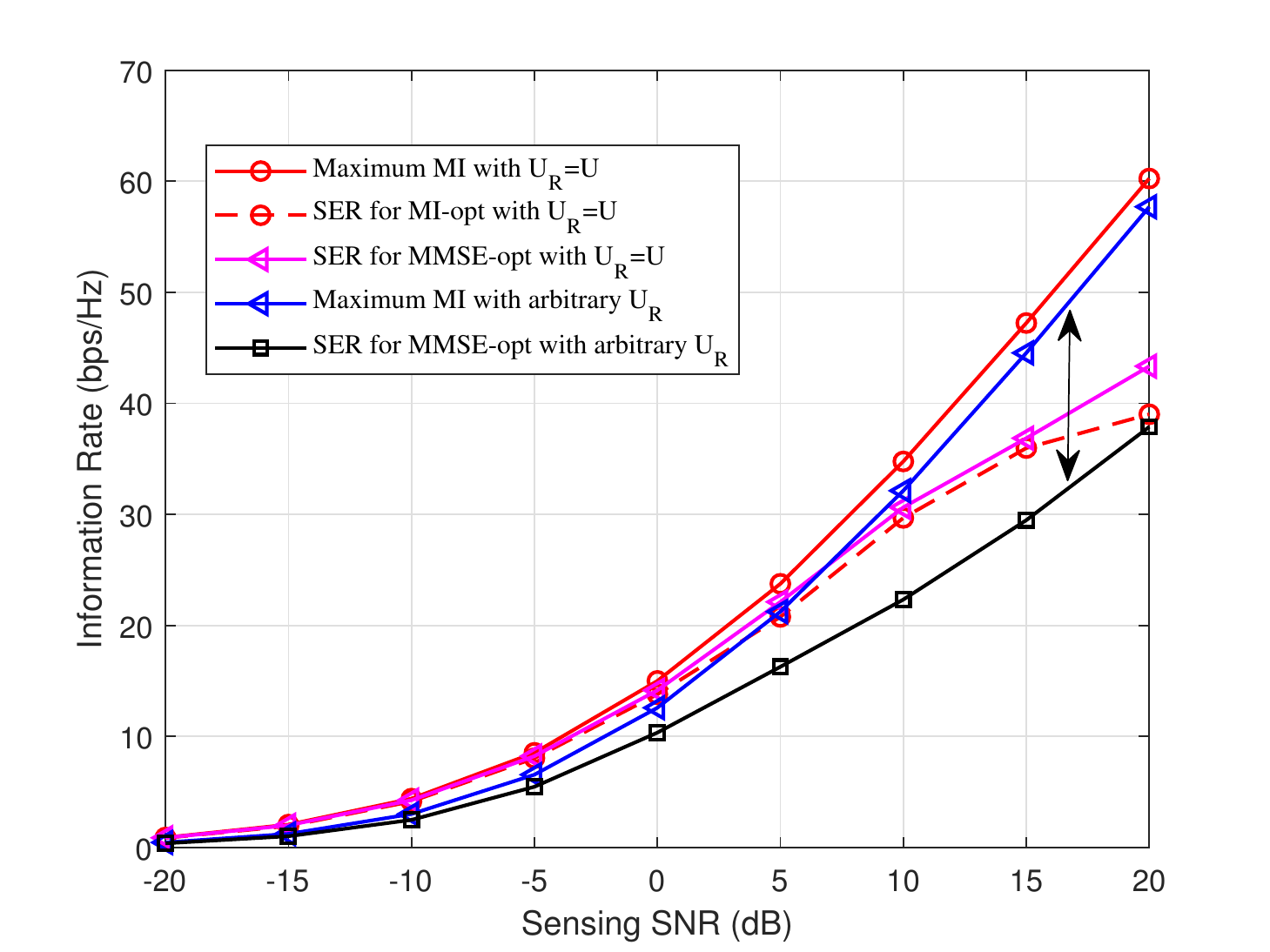}
	\caption{The influence of matrix $\textbf{F}$ in the semi-controllable GLM case.}
	\label{semicontrol}
\end{figure}

\section{Simulation Results}\label{simulation}

In this section, we provide the numerical simulations to validate the relationships between the MI, MMSE, and SER as theoretically derived in Section III. The parameters are set as $M=10$, $T=20$, $P_T=1$W. The entries of channel matrix $\textbf{H}$ are generated by standard complex Gaussian distribution.   

Fig. \ref{RandX} shows the curves of MI, MMSE, and SER as a function of sensing SNR in the GLM scenario. The following three phenomena can be observed: (1) The SER monotonically increases as the distortion $D=\text{MMSE}$ decreases. (2) The SER coincides with the MI when using the optimal waveform $\textbf{X}^\star$. (3) There exists a gap between SER and MI when employing an arbitrarily generated waveform.      

Fig. \ref{semicontrol} illustrates the influence of the matrix $\mathbf{F}$ in semi-controllable GLM case. The case where the equality holds is excluded as it is similar to the GLM. Here, we randomly generate the eigenvalues of matrix $\mathbf{F}$ while varying its right singular space. For $\textbf{U}_\text{R}=\textbf{U}$, SER for MI-opt waveform and SER for MMSE-opt waveform represent the SER corresponding to $D=\text{MMSE}(\mathbf{X}^\star)$ and $D=\text{MMSE}(\tilde{\mathbf{X}}^\star)$ respectively, where $\mathbf{X}^\star$ and $\tilde{\mathbf{X}}^\star$ denote the optimal solutions of maximizing MI and minimizing MMSE. We observe that $\mathbf{X}^\star \neq \tilde{\mathbf{X}}^\star $ leads to SER-MMSE $\ge$ SER-MI, and there exists a gap between SER-MMSE and maximum MI. For an arbitrary unitary $\textbf{U}_\text{R} \ne \textbf{U}$, we can see a significant reduction in the information rate compared to the first case,  which is consistent with the conclusion stated in Theorem 2.           

\section{Conclusion} 
This paper has proposed a novel efficiency metric for wireless sensing, namely, the sensing estimation rate (SER). Under the framework of the proposed SER, the sensing process can be analogized as an uncooperative lossy data transmission through joint source-channel coding. The bounds of SER and its achievability have been derived for the Gaussian linear model (GLM) and semi-controllable GLM. Finally, by comparing to the existing works, we show that the proposed framework may provide new information-theoretic insights into wireless sensing as well as optimal waveform design.

\begin{appendices}  
	\section{}\label{AppendixA} 
	\textit{Scalar Case}: The MMSE can be calculated by 
	\begin{equation}\label{MMSE_s} 
		D_L = \text{MMSE} = \frac{\sigma_{\upeta}^2}{1+|x|^2\sigma_{\upeta}^2/\sigma_z^2}=\frac{\sigma_{\upeta}^2}{1+\text{snr}},
	\end{equation} 
	where snr is defined by $|x|^2\sigma_{\upeta}^2/\sigma_z^2$. We have 
	\begin{equation} 
		R_E(D_L) = \log^+ \sigma_{\upeta}^2/D_L  = \log(1+\text{snr}) = C.
	\end{equation} 
	
	\textit{Vector Case}: In this paper, we assume the covariance matrix $\bm{\Sigma_{\bm{\upeta}}} \in \mathbb{C}^{M \times M}$ has full rank with $M=M_rM_t$ without loss of generality. The eigenvalue decomposition (EVD) of such a Hermitian matrix is defined by $\bm{\Sigma_{\bm{\upeta}}} = \textbf{U}\bm{\Lambda}\textbf{U}^H$. In what follows, we use the symbol $\bm{\Lambda}$ representing a diagonal matrix with $\bm{\Lambda}_i$ as its $i$-th eigenvalue, as well as $\textbf{U}$ representing a unitary matrix whose columns are the corresponding eigenvectors. By applying the EVD of matrix $\bm{\Sigma_{\bm{\upeta}}}$ with $\bm{\Lambda}_i=\sigma_{\upeta_i}^2$, the maximum MI for $I(\textbf{y};\bm{\upeta})$ is given by \cite{yang2007mimo}
	\begin{equation}\label{MaxMI}
		\begin{aligned}
			I(\textbf{X}^*) &= \mathop {\max} \limits_{\textbf{X} \in \mathcal{X}} \kern 2pt \log \left[\det (\sigma_z^{-2}\bm{\Sigma_{\bm{\upeta}}}\textbf{X}^H\textbf{X}+\textbf{I})\right] \\
			&=\sum_{i=1}^M\log \left[\left(\sigma_z^{-2}\sigma^2_{\upeta_i}\lambda-1\right)^+ +1\right],
		\end{aligned} 
	\end{equation}
	and the minimum MMSE counterpart can be expressed by
	\begin{equation}\label{MinMMSE}
		\begin{aligned}
			\text{MMSE}(\textbf{X}^*) &= \mathop {\min} \limits_{\textbf{X} \in \mathcal{X}} \kern 2pt \text{tr} \left\{ \left( \sigma_z^{-2} \textbf{X}^H\textbf{X} + \bm{\Sigma_{\bm{\upeta}}}^{-1} \right)^{-1} \right\} \\
			&=\sum_{i=1}^M \frac{\sigma^2_{\upeta_i}}{\left(\sigma_z^{-2}\sigma^2_{\upeta_i}\lambda-1\right)^+ +1},  \\
		\end{aligned} 
	\end{equation}
	where $\lambda$ is the water-filling factor such that $\sum_{i=1}^M (\lambda-\sigma_z^2/\sigma^2_{\upeta_i})^+ = TP_T$. In \cite{yang2007mimo}, the authors have proved that (\ref{MaxMI}) and (\ref{MinMMSE}) have the same solution with the form of 
	\begin{equation}\label{solution}
		\textbf{X}^* = \bm{\Phi} \left( \text{diag}\left[(\lambda-\frac{\sigma_z^2}{\sigma^2_{\upeta_1}})^+, \cdots, (\lambda-\frac{\sigma_z^2}{\sigma^2_{\upeta_M}})^+\right]  \right)^{\frac{1}{2}} \textbf{U}^H,   
	\end{equation}
	where $\bm{\Phi}$ is an arbitrary $M_rT \times M$ matrix with orthogonal columns. To further reveal the relationship between (\ref{MaxMI}) and (\ref{MinMMSE}) under the rate-distortion framework (\ref{VRD}), we let
	\begin{equation}\label{DDD}
		D_L=\text{MMSE}(\textbf{X}^*), \kern 5pt D_i=\frac{\sigma^2_{\upeta_i}}{\left(\sigma_z^{-2}\sigma^2_{\upeta_i}\lambda-1\right)^+ +1},
	\end{equation}
	and choose the inverse water-filling factor as $\xi=\sigma_z^2/\lambda$. Thus, it can be readily verified that      
	\begin{equation}
		R_E(D_L) = I(\textbf{X}^*) = C(\textbf{X}^*),
	\end{equation}   
	which completes the proof.
	
	\section{}\label{AppendixB}
	Here we omit the straightforward proof for the scalar case and focus on the vector case. Let $\textbf{R}_x \triangleq \textbf{X}^H\textbf{X}$ and the EVD of $\textbf{R}_x=\textbf{U}_x\bm{\Lambda}_x\textbf{U}^H_x$, then the optimal waveform can be readily reconstructed in the form of (\ref{solution}), once the matrices $\textbf{U}_x$ and $\bm{\Lambda}_x$ are determined. The maximum MI can be recast by
	\begin{equation}\label{FMaxMI}
		\begin{aligned}
			I(\textbf{X}^*) &= \mathop {\max} \limits_{\textbf{X} \in \mathcal{X}} \kern 2pt \log \left[\det(\sigma_z^{-2}\textbf{F}\bm{\Sigma_{\bm{\upeta}}}\textbf{F}^H\textbf{X}^H\textbf{X}+\textbf{I})\right]\\
			&=\sum_{i=1}^M\log \left[\left(\sigma_z^{-2}\bm{\Lambda}_{\text{F}_i}\lambda_s-1\right)^+ +1\right],
		\end{aligned}
	\end{equation} 
	through the EVD of $\textbf{F}\bm{\Sigma_{\bm{\upeta}}}\textbf{F}^H = \textbf{U}_\text{F}\bm{\Lambda}_{\text{F}}\textbf{U}_\text{F}^H$. In this case, we have  
	\begin{equation}\label{MI11}
	\textbf{U}_x=\textbf{U}_\text{F}, \quad  \bm{\Lambda}_{x_i}= (\lambda_s-\frac{\sigma_z^2}{\bm{\Lambda}_{\text{F}_i}})^+,
	\end{equation}
	where $\lambda_s$ is the water-filling factor satisfying the power budget.
	
	By recalling the EVD of $\bm{\Sigma_{\bm{\upeta}}}$, the MMSE is given by
	\begin{equation}\label{29}
		\begin{aligned}
			&\text{MMSE}(\textbf{X}^\star) = \mathop {\min} \limits_{\textbf{X} \in \mathcal{X}} \text{tr} \left\{ \left( \sigma_z^{-2} \textbf{F}^H\textbf{R}_x\textbf{F} + \bm{\Sigma_{\bm{\upeta}}}^{-1} \right)^{-1} \right\}\\ 
			&= \mathop {\min} \limits_{\textbf{X} \in \mathcal{X}}\text{tr} \left\{ \left( \sigma_z^{-2} \bm{\Lambda}^{\frac{1}{2}}\textbf{U}^H\textbf{F}^H\textbf{R}_x\textbf{F}\textbf{U}\bm{\Lambda}^{\frac{1}{2}} + \textbf{I} \right)^{-1} \bm{\Lambda} \right\} \\
			&\mathop = \limits^{(d)}\mathop {\min} \limits_{\textbf{X} \in \mathcal{X}} \text{tr} \left\{ \left( \sigma_z^{-2}\textbf{U}_\text{R} \tilde{\bm{\Lambda}}^H_\text{F}\textbf{U}^H_\text{F}\textbf{R}_x\textbf{U}_\text{F}\tilde{\bm{\Lambda}}_\text{F}\textbf{U}_\text{R}^H + \textbf{I} \right)^{-1} \bm{\Lambda} \right\} \\
			&\mathop \ge \limits^{(e)} \mathop {\min} \limits_{\textbf{X} \in \mathcal{X}} \sum_{i=1}^M \frac{\sigma^2_{\upeta_i}}{\sigma_z^{-2}\bm{\Lambda}_{\text{F}_{i}}\bm{\Lambda}_{x_i} +1} \triangleq f(\bm{\Lambda}_{x}). 
		\end{aligned} 
	\end{equation}
	Here, $(d)$ holds due to the singular value decomposition (SVD) of $\textbf{F}\textbf{U}\bm{\Lambda}^{\frac{1}{2}}=\textbf{U}_\text{F}\tilde{\bm{\Lambda}}_\text{F}\textbf{U}_\text{R}^H$, where $\textbf{U}_\text{R}$ is the right singular matrix and we have
	\begin{equation}
	\textbf{F}\bm{\Sigma_{\bm{\upeta}}}\textbf{F}^H=\textbf{F}\textbf{U}\bm{\Lambda}^{\frac{1}{2}}\bm{\Lambda}^{\frac{1}{2}}\textbf{U}^H\textbf{F}^H=\textbf{U}_\text{F}\tilde{\bm{\Lambda}}_\text{F}\tilde{\bm{\Lambda}}_\text{F}^H\textbf{U}_\text{F}^H,
	\end{equation}   
	with $\tilde{\bm{\Lambda}}_\text{F}\tilde{\bm{\Lambda}}^H_\text{F}=\bm{\Lambda}_{\text{F}}$. Moreover, $(e)$ is due to the following lemma.

	\textit{Lemma 1}: For a positive definite Hermitian matrix $\textbf{A} \in \mathbb{C}^{M \times M}$ and a diagonal matrix $\bm{\Lambda}$ with its entries are positive, then the following inequality holds:
	\begin{equation} 
		\text{tr}({\textbf{A}^{-1}\bm{\Lambda}}) \ge \sum_{i=1}^{M} \frac{\bm{\Lambda}_{ii}}{\textbf{A}_{ii}},
	\end{equation}     
	the equality holds if and only if $\textbf{A}$ is diagonal. The proof can be readily obtained through the similar procedure in \cite[Appendix I]{ohno2004capacity}. Based on \textit{Lemma 1}, the equality of $(e)$ holds if and only if $\textbf{U}_x=\textbf{U}_\text{F}$ and $\textbf{U}_\text{R}$ is an identity matrix.  
	 	
	Furthermore, the optimal solution of $f(\bm{\Lambda}_{x})$ is given by 
	\begin{equation}\label{MM11}
	\bm{\Lambda}_{x_i}^\star = \left(\sqrt{\frac{\sigma_z^2\sigma^2_{\upeta_i}}{\bm{\Lambda}_{\text{F}_{i}}\tilde{\lambda}_m}}-\frac{\sigma_z^2}{\bm{\Lambda}_{\text{F}_{i}}}\right)^+ \triangleq \left(a_i\lambda_m-\frac{\sigma_z^2}{\bm{\Lambda}_{\text{F}_{i}}}\right)^+, 
	\end{equation}  
	where $\lambda_m$ is the water-filling factor. By comparing (\ref{MM11}) and (\ref{MI11}), we can observe that the solutions of maximizing MI and minimizing MMSE are no longer the same even if the equality of $(e)$ holds. This is because the weight factor $a_i$ may change the relative states among the parallel sub-channels, thereby leading to different water-filling solutions. Therefore, $f(\bm{\Lambda}_{x})$ and $I(\textbf{X})$ have the same solution when $a_i = a$ remains constant for each $i$. Namely, we have $\bm{\Lambda}_{\text{F}_{i}}=a\sigma^2_{\upeta_i}$, which implies matrix $\textbf{F}$ has identical non-zero eigenvalues.
	
	Consequently, according to the monotonicity of the rate-distortion function and (\ref{VRD}), we have
	\begin{equation}
		R_E(\text{MMSE}(\textbf{X}^\star)) \le R_E(f(\bm{\Lambda}_{x}^\star)) = I(\bm{\Lambda}_{x}^\star)  \le I(\textbf{X}^*).
	\end{equation}  
	The first equality holds if and only if $\textbf{U}_\text{R}$ is an identity matrix, which means that the right singular space of $\textbf{F}$ equals to the eigenspace of $\bm{\Sigma_{\bm{\upeta}}}$. The last equality holds when $\textbf{F}$ has identical non-zero eigenvalues. This completes the proof.    
		
	\section{}\label{AppendixC}
	The Bayesian Fisher information matrix (BFIM) of $\bm{\upeta}$ conditioned on $\textbf{X}$ can be expressed by
	\begin{equation}
		\begin{aligned}
			\textbf{J}_{\bm{\upeta}|\textbf{X}}&= \textbf{J}_{\text{D}} + \textbf{J}_\text{P}=\mathbb{E}_{\bm{\upeta}}\left[ \sigma^{-2}_z \textbf{F}^H_{\bm{\upeta}} \textbf{X}^H \textbf{X} \textbf{F}_{\bm{\upeta}}  \right]+ \textbf{J}_\text{P} \\
			&\mathop = \limits^{(f)} \sigma^{-2}_z \textbf{G}^H \textbf{X}^H \textbf{X} \textbf{G} + \textbf{J}_\text{P},
		\end{aligned}
	\end{equation} 
	where $\textbf{F}_{\bm{\upeta}} = \partial \textbf{h}(\bm{\upeta})/\partial \bm{\upeta}$ is the Jacobin matrix. $\textbf{J}_\text{P}$ is the prior Fisher information provided by the prior distribution $p_{\bm{\upeta}}(\bm{\eta})$, which can be obtained by
	\begin{equation}
		\textbf{J}_\text{P} = \mathbb{E}_{\bm{\upeta}} \left[ \frac{\partial \ln p_{\bm{\upeta}}(\bm{\eta})}{\partial \bm{\upeta}} \frac{\partial \ln p_{\bm{\upeta}}(\bm{\eta})}{\partial \bm{\upeta}^T}   \right].
	\end{equation} 
	$(f)$ follows the results of \cite[Proposition 2]{xiong2022flowing}, where $\textbf{G}$ can be expressed by 
	\begin{equation}
		\textbf{G}=\sqrt{\lambda_\Psi}\text{mat}_{M \times K}(\textbf{u}),
	\end{equation}   
	where $\text{mat}_{M \times K} (\textbf{u})$ denotes the $M \times K$ matrix satisfying $\text{vec}(\text{mat}_{M \times K} (\textbf{u}))=\textbf{u}$, and
	\begin{equation}
		\bm{\Psi}=\mathbb{E}_{\bm{\upeta}} \left[{\text{vec}(\textbf{F}_{\bm{\upeta}})\text{vec}(\textbf{F}_{\bm{\upeta}})^H}\right]=\lambda_\Psi\textbf{u}\textbf{u}^H.
	\end{equation}
	The rightmost equality is the EVD of matrix $\bm{\Psi}$. The above process follows the principle of Choi representation theory, whose details can be found in \cite{xiong2022flowing}. Therefore, the minimum BCRB is   
	\begin{equation}
		\text{BCRB}(\textbf{X}^\star)=\mathop {\min} \limits_{\textbf{X} \in \mathcal{X}} \kern 2pt \text{tr} \left\{ \left( \sigma_z^{-2} \textbf{G}^H \textbf{R}_x\textbf{G} + \textbf{J}_\text{P} \right)^{-1} \right\},
	\end{equation}
	which has the similar form as (\ref{29}) in semi-controllable GLM scenario. Following the same procedure in Appendix \ref{AppendixB}, we have 
	\begin{equation}
		R_E(\text{BCRB}(\textbf{X}^\star)) \le \sum_{i=1}^M\log \left(\sqrt{\frac{\sigma_z^2\bm{\Lambda}_{\text{J}_{i}}}{\bm{\Lambda}_{\text{G}_{i}}}}\lambda-\frac{\sigma_z^2}{\bm{\Lambda}_{\text{G}_{i}}}\right)^+,
	\end{equation}
	where $\lambda$ is the water filling factor, $\bm{\Lambda}_{\text{J}}$ and $\bm{\Lambda}_{\text{G}}$ are the diagonal matrices satisfying the EVDs of $\textbf{J}_\text{P}^{-1}=\textbf{U}_\text{J}\bm{\Lambda}_{\text{J}}\textbf{U}_\text{J}^H$ and $\textbf{G}\textbf{J}_{\text{P}}\textbf{G}^H = \textbf{U}_\text{G}\bm{\Lambda}_{\text{G}}\textbf{U}_\text{G}^H$. The equality holds if and only if the right singular space of $\textbf{G}$ equals to $\textbf{U}_\text{J}$, which completes the proof.
	
\end{appendices} 
      
\bibliographystyle{IEEEtran}
\bibliography{IEEEabrv,SER_TVT_Revised}

\begin{thebibliography}{10}
\providecommand{\url}[1]{#1}
\csname url@samestyle\endcsname
\providecommand{\newblock}{\relax}
\providecommand{\bibinfo}[2]{#2}
\providecommand{\BIBentrySTDinterwordspacing}{\spaceskip=0pt\relax}
\providecommand{\BIBentryALTinterwordstretchfactor}{4}
\providecommand{\BIBentryALTinterwordspacing}{\spaceskip=\fontdimen2\font plus
\BIBentryALTinterwordstretchfactor\fontdimen3\font minus
  \fontdimen4\font\relax}
\providecommand{\BIBforeignlanguage}[2]{{%
\expandafter\ifx\csname l@#1\endcsname\relax
\typeout{** WARNING: IEEEtran.bst: No hyphenation pattern has been}%
\typeout{** loaded for the language `#1'. Using the pattern for}%
\typeout{** the default language instead.}%
\else
\language=\csname l@#1\endcsname
\fi
#2}}
\providecommand{\BIBdecl}{\relax}
\BIBdecl

\bibitem{liu2022integrated}
F.~Liu, Y.~Cui, C.~Masouros, J.~Xu, T.~X. Han, Y.~C. Eldar, and S.~Buzzi,
  ``Integrated sensing and communications: Towards dual-functional wireless
  networks for 6g and beyond,'' \emph{IEEE journal on selected areas in
  communications}, 2022.

\bibitem{liu2022survey}
A.~Liu \emph{et~al.}, ``A survey on fundamental limits of integrated sensing
  and communication,'' \emph{IEEE Communications Surveys \& Tutorials},
  vol.~24, no.~2, pp. 994--1034, 2022.

\bibitem{thomas2006elements}
M.~Thomas and A.~T. Joy, \emph{Elements of information theory}.\hskip 1em plus
  0.5em minus 0.4em\relax Wiley-Interscience, 2006.

\bibitem{bell1993information}
M.~R. Bell, ``Information theory and radar waveform design,'' \emph{IEEE
  Transactions on Information Theory}, vol.~39, no.~5, pp. 1578--1597, 1993.

\bibitem{tang2018spectrally}
B.~Tang and J.~Li, ``Spectrally constrained {MIMO} radar waveform design based
  on mutual information,'' \emph{IEEE Transactions on Signal Processing},
  vol.~67, no.~3, pp. 821--834, 2018.

\bibitem{yang2007mimo}
Y.~Yang and R.~S. Blum, ``{MIMO} radar waveform design based on mutual
  information and minimum mean-square error estimation,'' \emph{IEEE
  Transactions on Aerospace and Electronic Systems}, vol.~43, no.~1, pp.
  330--343, 2007.

\bibitem{ahmadipour2022information}
M.~Ahmadipour, M.~Kobayashi, M.~Wigger, and G.~Caire, ``An
  information-theoretic approach to joint sensing and communication,''
  \emph{IEEE Transactions on Information Theory}, 2022.

\bibitem{chiriyath2015inner}
A.~R. Chiriyath, B.~Paul, G.~M. Jacyna, and D.~W. Bliss, ``Inner bounds on
  performance of radar and communications co-existence,'' \emph{IEEE
  Transactions on Signal Processing}, vol.~64, no.~2, pp. 464--474, 2015.

\bibitem{li2007range}
J.~Li, L.~Xu, P.~Stoica, K.~W. Forsythe, and D.~W. Bliss, ``Range compression
  and waveform optimization for mimo radar: A cram{\'e}r--rao bound based
  study,'' \emph{IEEE Transactions on Signal Processing}, vol.~56, no.~1, pp.
  218--232, 2008.

\bibitem{4600229}
M.~Zhao, Z.~Shi, and M.~C. Reed, ``Iterative turbo channel estimation for ofdm
  system over rapid dispersive fading channel,'' \emph{IEEE Transactions on
  Wireless Communications}, vol.~7, no.~8, pp. 3174--3184, 2008.

\bibitem{liu2022deterministic}
F.~Liu, Y.~Xiong, K.~Wan, T.~Han, and G.~Caire, ``Deterministic-random tradeoff
  of integrated sensing and communications in {G}aussian channels: A
  rate-distortion perspective,'' \emph{accepted in IEEE International Symposium
  on Information Theory (ISIT) 2023, arXiv:2212.10897}.

\bibitem{yp2023information}
Y.~Polyanskiy and Y.~Wu, \emph{Information Theory: From Coding to Lerning},
  1st~ed.\hskip 1em plus 0.5em minus 0.4em\relax Cambridge University Press,
  2023.

\bibitem{liu2020radar}
F.~Liu, W.~Yuan, C.~Masouros, and J.~Yuan, ``Radar-assisted predictive
  beamforming for vehicular links: Communication served by sensing,''
  \emph{IEEE Transactions on Wireless Communications}, vol.~19, no.~11, pp.
  7704--7719, 2020.

\bibitem{ohno2004capacity}
S.~Ohno and G.~B. Giannakis, ``Capacity maximizing mmse-optimal pilots for
  wireless {OFDM} over frequency-selective block rayleigh-fading channels,''
  \emph{IEEE Transactions on Information Theory}, vol.~50, no.~9, pp.
  2138--2145, 2004.

\bibitem{xiong2022flowing}
Y.~Xiong, F.~Liu, Y.~Cui, W.~Yuan, and T.~X. Han, ``On the fundamental tradeoff
  of integrated sensing and communications under gaussian channels,''
  \emph{IEEE Transactions on Information Theory}, Accepted, {DoI}:
  10.1109/TIT.2023.3284449, 2023.

\end{thebibliography}

\end{document}